\documentclass[%
 reprint,
 amsmath,amssymb,
 aps
]{revtex4-1}

\usepackage{amsmath}
\usepackage{amssymb}
\usepackage{graphicx}
\usepackage{dcolumn}
\usepackage{bm}
\usepackage{makecell}

\newcommand{\abs}[1]{\left\vert #1 \right\vert}

\begin{document}

\preprint{APS/123-QED}

\title{Phase Behavior of Charged Colloids at a
  Fluid Interface}

\author{Colm~P.~Kelleher}
\affiliation{Department of Physics and Center for Soft Matter Research, New York University, 4 Washington Place, New York, New York 10003, USA}
\author{Rodrigo E. Guerra}
\affiliation{Department of Physics and Center for Soft Matter Research, New York University, 4 Washington Place, New York, New York 10003, USA}
\author{Andrew D. Hollingsworth}
\affiliation{Department of Physics and Center for Soft Matter Research, New York University, 4 Washington Place, New York, New York 10003, USA}
\author{Paul~M.~Chaikin}
\affiliation{Department of Physics and Center for Soft Matter Research, New York University, 4 Washington Place, New York, New York 10003, USA}

\date{\today}

\begin{abstract} 
We study the phase behavior of a system of charged
colloidal particles that are electrostatically bound to an almost flat
interface between two fluids. We show that, despite the fact that our
experimental system consists of only $10^{3}$ - $10^{4}$ particles,
the phase behavior is consistent with the theory of melting due to
Kosterlitz, Thouless, Halperin, Nelson and Young (KTHNY). Using spatial
and temporal correlations of the bond-orientational order parameter,
we classify our samples into solid, isotropic fluid, and hexatic
phases. We demonstrate that the topological defect structure we
observe in each phase corresponds to the predictions of KTHNY
theory. By measuring the dynamic Lindemann parameter,
$\gamma_{L}(\tau)$, and the non-Gaussian parameter, $\alpha_{2}(\tau)$,
of the displacements of the particles relative to their neighbors, we
show that each of the phases displays distinctive dynamical behavior.
\end{abstract}

\maketitle

\section{Introduction} 
Colloidal systems have long been used as a model for investigating
fundamental questions in condensed matter physics. Two dimensional
(2D) systems are of particular interest, both for the rich physical phenomena they display
\cite{MurrayPRB1990,ZahnPRL1999} and for the ease with which
they can be imaged, via video or confocal microscopy
\cite{CrockerGrier1996}. To create such a system, colloidal
particles must somehow be confined to a surface. This can be done by
using colloids sedimented onto a solid or fluid substrate
\cite{ZahnPRL1999,BubeckPRL1999,SkinnerPRL2010}, by physically
confining colloids between the parallel walls of a thin sample chamber
\cite{MurrayPRB1990,ZhengPRL2011}, or by using charged particles that
bind electrostatically to a fluid interface \cite{Leunissen2007,
IrvinePNAS2013}. The latter system has the advantage that the surface
to which the particles bind does not have to be flat, and so is
particularly useful in exploring the role of background curvature in
determining the structure and dynamics of topological defects in 2D
materials \cite{IrvineNature2010, IrvineNatMat2012}. However, the
phase behavior of colloids in this kind of system, which is necessary
for a full understanding of experiments undertaken at finite
temperature, has not been investigated.

\begin{figure}[t!]
\begin{centering}
\includegraphics[scale=0.23]{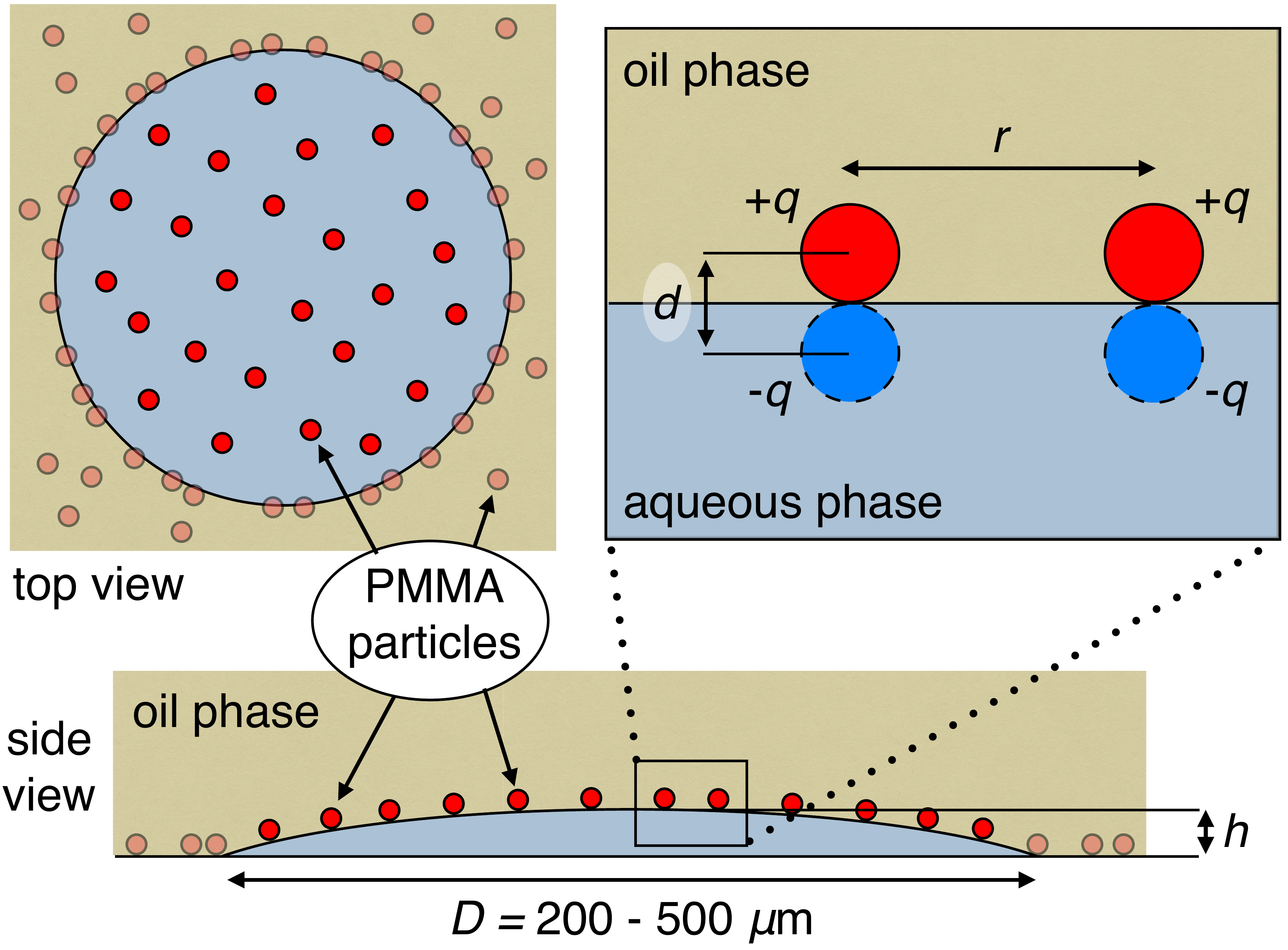}
\caption{(Color online.) Schematic of the experimental geometry. In
the ``side view'' panel, the height $h$ of the droplet is exaggerated
compared to the base diameter $D$. In experimental samples, $h\lesssim
5 \,\mu$m. Particles sitting on the interface (shown in full color)
are mobile, while the remaining particles (shown as faded) bind
randomly and irreversibly to the bare glass surface. The boundary of
the droplet is delineated by a row of particles stuck to the
glass. The inset shows the origin of the dipolar repulsion between
interfacial particles.
\label{fig:conf_diag}}
\end{centering}
\end{figure}

In this work, we study systems of 10$^{3}$ - 10$^{4}$ charged
colloidal particles that are electrostatically bound to an almost-flat
fluid interface, shown schematically in Fig.~\ref{fig:conf_diag}. We
demonstrate that the interaction between particles is consistent with
a dipolar pair potential. We measure the dipole moment of the
particles, which allows us to directly compare the phase behavior of
our system to previous experiments and simulations using dipolar
particles~\cite{LinPRE2006,ZahnPRL1999, DeutschlanderPRL2014}.

Using density as the control parameter, we show that the phase behavior of our
system is consistent with the theory of defect-mediated melting due to
Kosterlitz, Thouless, Halperin, Nelson and Young (KTHNY), whereby the
transition from isotropic fluid to crystalline solid happens via an
intermediate hexatic phase \cite{KosterlitzThoulessJPhys1973,
YoungPRB1979, HalperinNelsonPRL1979}. We identify the solid, isotropic
fluid, and hexatic phases by measuring the bond-orientational order
parameter $\psi_{6}$ and associated space and time correlation
functions $g_{6}(r)$ and $g_{6}(\tau)$. Finally, we show that the
classification of our samples into solid, isotropic fluid and hexatic phases is
consistent both with the topological defect structure predicted by
KTHNY theory, and with the dynamical behavior that has been described
previously \cite{ZahnPRL2000}.

\begin{figure*}[t!]
\center
\includegraphics[scale=0.36]{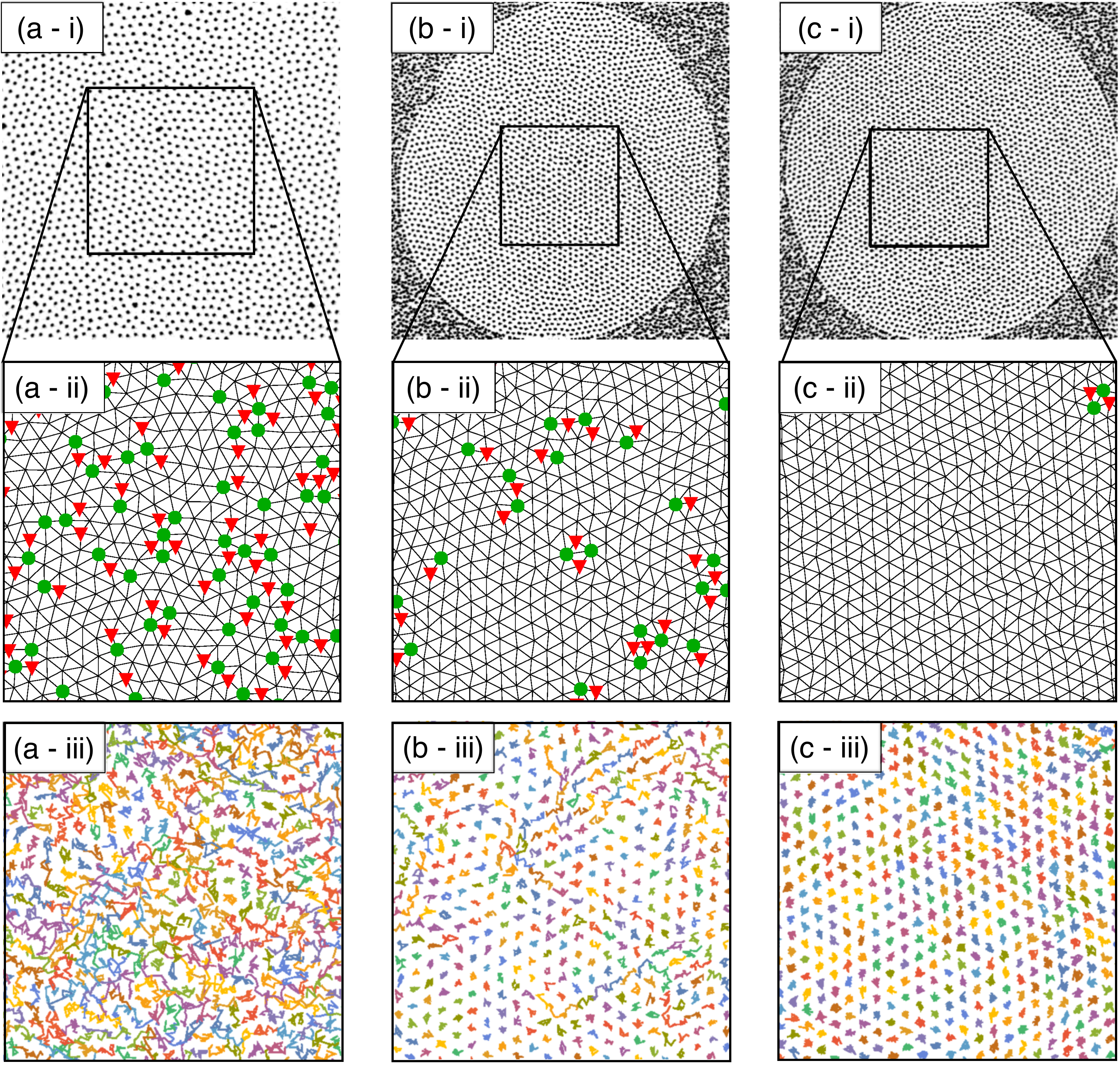}
\caption{(Color online.) Three samples of hydrophobic PMMA particles
electrostatically bound to almost-flat droplets, at areal densities
$\rho$ of (a) $0.036$, (b) $0.043$, and (c) $0.049 \,\mu$m$^{-2}$.
The top row shows a confocal micrograph of each sample. For samples
(b) and (c), the entire, roughly circular droplet is shown (see
Fig.~\ref{fig:conf_diag}). In each sample, the inset square has a side
length of 100$\,\mu$m. The middle row shows Delaunay triangulations of
the particle positions in a selected region, in the first frame of
each movie. Particles with five or seven nearest neighbors are respectively
indicated by red triangles and green disks. The bottom row shows
the particle trajectories over 25 minutes.
\label{fig:confocal_plus_trajectoires}}
\end{figure*}

\section{\label{sec:Materials}Materials \& methods} 
Our experimental system is composed of diameter $d=1.1 \, \mu$m
poly(methyl methacrylate) (PMMA) particles that bind electrostatically
to the interface between an oil and an aqueous phase.  The particles
are initially dispersed in a 1:1 volumetric mixture of cyclohexyl
bromide (CHB) and dodecane, while the aqueous phase consists of a 10mM
solution of NaCl in a 90 wt\% glycerol-water mixture.  The dielectric
constant $\epsilon$ of the oil is $4.3\, \epsilon_{0}$, where
$\epsilon_{0}$ is the permittivity of the vacuum
\cite{YoolengaPhysica1965}. The CHB is purified and stored according
to the protocols given in references \cite{Leunissen2007} and \cite{
EvansPRE2016}. The PMMA particles are sterically stabilized with
covalently bound poly(12-hydroxystearic acid) (PHS)
\cite{ElsesserHollingsworth2010}. Previous work has shown that, when
dispersed in similar oils, micron-sized PHS-coated PMMA particles
acquire a charge $q$ of around +500$\,$e, where e is the elementary
charge \cite{KelleherPRE2015, Leunissen2007}. While the charging
mechanism is still incompletely understood \cite{MirjamLeunissenPhD,
vanderLindenLangmuir2014, EvansPRE2016}, particle
charging in our system is robust and reproducible \cite{KelleherPRE2015}. To
facilitate measurement of particle dynamics with confocal microscopy,
we fluorescently dye the particles with absorbed rhodamine 6G
\cite{ElsesserEtAl2011}.

To prepare the particle-laden interfaces we use in our experiments, we
use an atomizer to deposit droplets of the aqueous phase onto a cover
slip. We then incorporate the cover slip into the construction of a
glass capillary channel, which is filled with the particle dispersion
at the desired concentration. As the particle dispersion flows into
the chamber, some of the particles bind irreversibly to the surface of
the droplets, while others bind to the bare glass surface. The
experimental geometry is shown schematically in
Fig.~\ref{fig:conf_diag}. Particles that are bound to the interface
are mobile, and can reach thermodynamic equilibrium. To control the
flatness of the fluid interfaces, prior to the droplet deposition
step, the cover slip is immersed in a bath of KOH-saturated
isopropanol (IPA), and rinsed sequentially with DI water, acetone and
IPA. The cover slip is blown dry with an N$_{2}$ sprayer and dried in
an oven at 70$^{\circ}$C for at least 15 mins prior to use. By varying
the immersion time of the cover slips in the KOH solution, we control
the advancing contact angle of the deposited droplets of the aqueous
phase \cite{Note1}. We find that
immersing the cover slips for 30 mins.~gives contact angles of
approximately 1$^{\circ}$, which are appropriate for this
experiment. Apart from the KOH immersion step, we follow the same
protocol to clean all glass surfaces that come into contact with the
particle dispersion. Once the capillary channel has been filled, we
seal it: first with a buffer layer of glycerol and then with optical
adhesive (Norland Products Inc. NOA \#68).

Following the above procedures, we obtain a sample chamber that
contains several particle-laden interfaces, ranging in base diameter,
$D$, from around 200 to 500$\,\mu$m. Each interface has a slightly
different areal density, $\rho$, of PMMA particles, thus allowing us
to approximately uniformly sample areal densities in the range
0.01-0.15$\,\mu$m$^{-2}$. We estimate the
curvature of the droplets as follows: using a 10$\times$ magnification
NA 0.3 air objective mounted on a Leica TCS SP5 II confocal
microscope, we image the particles in a single confocal slice.  If all
the particles appear in the field of view, the maximum thickness of
the drop must then be less than the the optical section thickness,
around $5\,\mu$m. Since the dimensions of our droplet are far smaller
than the capillary length, we ignore the effect of gravity
\cite{Note2}, and assume that the droplets take the shape of a spherical cap. Since
$D >200\,\mu $m, and the thickness is $<5\,\mu$m, the radius of
curvature must be at least $1\,$mm: far greater than the length scales
probed in our experiments.  Thus, when analyzing our experimental
data, we treat the droplet surface as flat.

After waiting at least a day for the samples to equilibriate, we use
confocal microscopy to record the motion of the particles for up to
several hours, at a rate of 0.25-1.0 frames/s.  Using standard
routines \cite{CrockerWeeksTracking}, we locate the
particles in the field of view. Delaunay triangulations of the
instantaneous particle positions identify sites with more or fewer
than six nearest neighbors, called disclinations.  Trajectories
obtained by by linking particle positions in adjacent frames reveal
the mobility of individual particles.

Snapshots and movies of particle layers of different densities
captured and analyzed in this way display the qualitative features of
the three phases predicted by KTHNY.  The low-density, isotropic fluid
phase shown in Fig.~\ref{fig:confocal_plus_trajectoires} (a) is
characterized by homogeneously distributed disclination defects and
uniformly mobile particles. By contrast, the high-density, equilibrium
crystal phase, shown in Fig.~\ref{fig:confocal_plus_trajectoires} (c),
is only capable of supporting sparse clusters of tighly-bound defects
that do not affect the long-range order of the lattice. In this phase, the
particles that compose the crystal are uniformly confined to the
vicinity of their lattice sites.  As we show in Section~\ref{sec:model},
the resulting caged diffusion can be used to measure in-situ the strength of
the inter-particle interactions.

At intermediate particle densities, such as that shown in
Fig.~\ref{fig:confocal_plus_trajectoires} (b), isolated disclinations
condense into the defect clusters that characterize the hexatic phase.
These clusters facilitate particle mobility, disrupt translational
order, and induce long-lived spatial inhomogeneities in the structure
and dynamics of the particles.

\begin{figure}[t!]
\center
\includegraphics[scale=0.45]{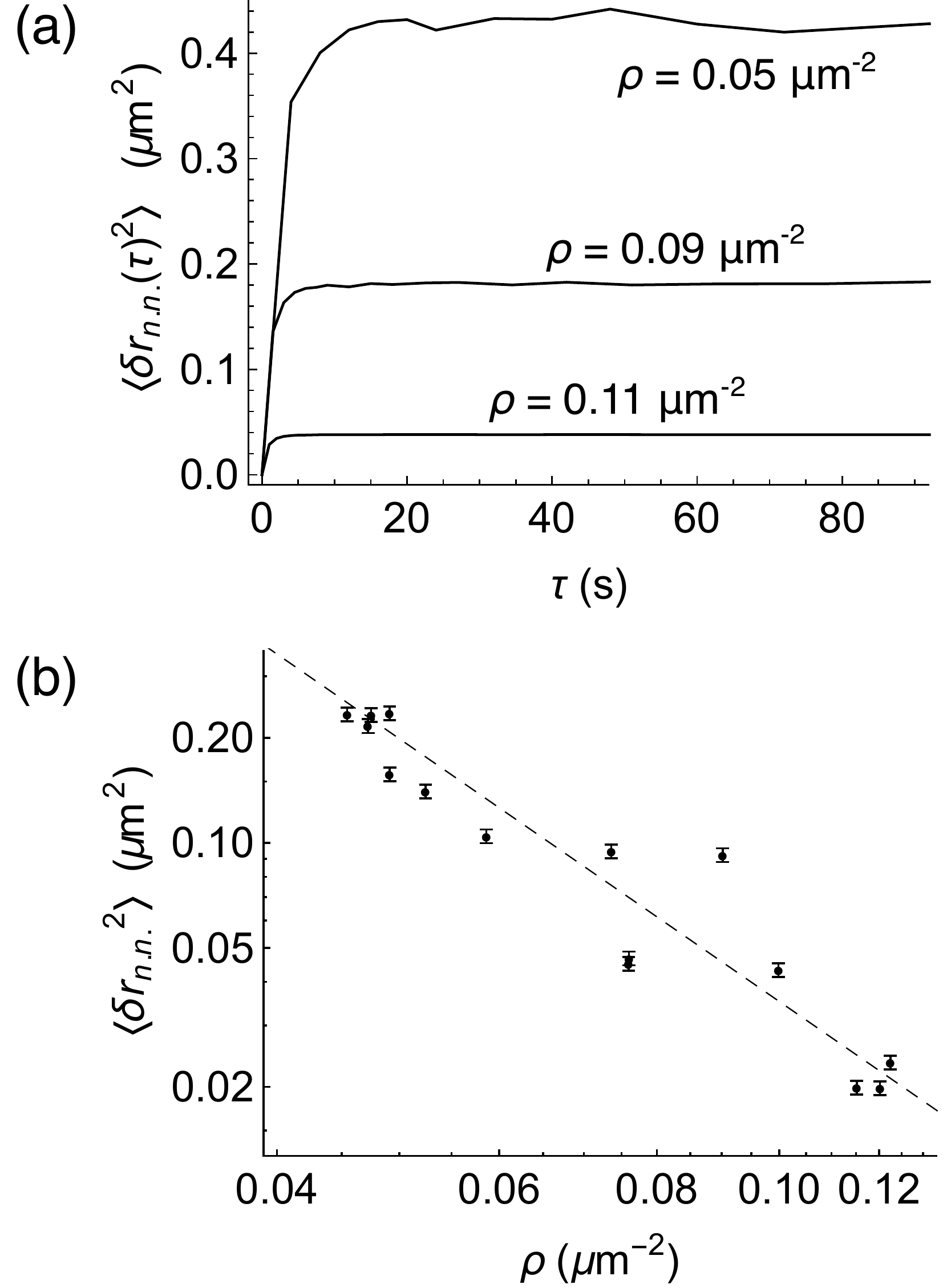}
\caption{(a) Nearest-neighbor relative mean square displacement curves for
three crystalline samples, at different densities $\rho$. (b)
Log-log plot of the limiting value of the n.n.-MSD, as a function of
areal density. The dashed line is the best-fit line with
slope -5/2, as predicted by Eqn.~\eqref{eq:drVsRho}. This fit
gives an electric dipole moment $p = (455 \pm 20)\,$e$\cdot \mu$m.
\label{fig:dr2VsRho}}
\end{figure}

\section{\label{sec:model}Model for interparticle interactions \& measurement of
  dipole moment of particles}
To compare the phase behavior of our system with that observed in
other systems~\cite{LinPRE2006, ZahnPRL1999, HanPRE2008,
DeutschlanderPRL2014}, it is important to know both the form and
magnitude of the interparticle repulsion~\cite{KapferKrauthPRL2015}.
In previous work, we studied the behavior of PHS-coated PMMA particles
at an interface between two fluids which are similar in composition to
those described here~\cite{KelleherPRE2015}. In our system, the PMMA
particles appear to be wet very little (or possibly not at all) by the
aqueous phase, and can be described as spheres with charge $q$ sitting
on top of a conducting medium, as shown in the inset to
Fig.~\ref{fig:conf_diag}. In the same work, we showed that the force
binding individual particles to the interface is electrostatic in
origin, and that the interaction between pairs of particles is
dipolar.  These results are consistent with previous experimental and
theoretical work \cite{GouldingHansenMP1998, AveyardPRL2002,
OettelPRE2007} on systems of charged particles in the vicinity of a
fluid interface. In our system and similar ones, the interaction of
two charged particles is the sum of the Coulomb repulsion between the
particles and the Coulomb attraction between each particle and the
image charge of the other. Thus, in the limit where the interparticle
distance $r$ is large compared to the diameter $d$ of the particles,
the net interaction between two interfacially bound particles can be
approximated by a pair potential of the form:
\begin{equation*} 
U(r) \simeq \frac{A}{r^{3}},
\end{equation*} 
where $ A =p^{2} /8 \pi \epsilon$, and $p = qd$ is the magnitude of
the electric dipole moment of the particles \cite{Note3}.  As shown in
Fig.~\ref{fig:confocal_plus_trajectoires} (c-iii), when the
interfacial density of the particles is high enough, they form a
crystalline solid phase. In a hexagonal lattice composed of repulsive
dipolar particles, the average value of $p$ can be estimated by
observing the fluctuations of the particles relative to the cage
formed by their nearest neighbors. To quantify the interactions
between our interfacial PMMA particles, we use a method due to
Parolini et al.~\cite{ParoliniJCPM2015}, which we now outline briefly.

We begin by identifying a region of the interface where several
hundred particles are arranged in a defect-free crystal lattice which
is at least 20$\,\mu$m from the droplet boundary. We consider only
samples where the density gradient across the subregion of interest is
less than 0.3\% per interparticle spacing, and avoid the grain
boundaries or isolated dislocations that are occasionally present in
our samples. These non-equilibrium features may be identified
quantitatively, for instance by anomalous behavior of the dynamic
Lindemann parameter, $\gamma_{L}(\tau)$, or the non-Gaussian parameter,
$\alpha_{2}(\tau)$, which we define in Section~\ref{sec:dynamics}. To
further check that we are measuring equilibrium properties, we verify
that our results do not depend strongly on the particular choice of
subregion.

At each instant $t$ in time, the Delaunay triangulation defines
$\mathcal{N}_{i}$, the set of nearest neighbors of particle $i$. The
position of particle $i$ relative to its neighbors is given by
\begin{equation*} 
\mathbf{r}_{i, \text{n.n.}}(t) =\mathbf{r}_{i}(t) - \frac{1}{n_{b}}
\sum_{j  \, \in \, \mathcal{N}_{i}}\mathbf{r}_{j}(t),
\end{equation*} 
where the sum is taken over the $n_{b}$ neighbors of particle $i$.
The nearest-neighbor relative mean square displacement (n.n.-MSD) as a
function of time interval $\tau$ is defined as
\begin{equation} 
\langle \delta \mathbf{r}_{\text{n.n.}}(\tau)^{2} \rangle =
\langle (\mathbf{r}_{i, \text{n.n.}}(t+\tau) - \mathbf{r}_{i,
  \text{n.n.}}(t))^{2}\rangle,
\label{eq:nnMSd}
\end{equation} 
where the average is taken over particles $i$ and starting times
$t$. When calculating the quantity $\mathbf{r}_{i,
\text{n.n.}}(t+\tau)$, we use the set of neighbors defined at time
$t$, even if those particles no longer share a Delaunay bond with
particle $i$ at time $t+\tau$. The upper panel of
Fig.~\ref{fig:dr2VsRho} shows $\langle \delta
\mathbf{r}_{\text{n.n.}}(\tau)^{2} \rangle$ curves three crystalline
samples, at different densities $\rho$. In all these samples, $\langle
\delta \mathbf{r}_{\text{n.n.}}(\tau)^{2} \rangle$ reaches a plateau
value $\langle \delta \mathbf{r}_{\text{n.n.}}^{2}\rangle$. According
to~\cite{ParoliniJCPM2015}, this value is related to the force
constant $A$, and hence to the dipole moment $p$, by the equation
\begin{equation} 
\langle \delta \mathbf{r}_{\text{n.n.}}^{2} \rangle = \frac{2^{9/2}
\alpha \, k_{B}T}{3^{5/4}A}\rho^{-5/2},
\label{eq:drVsRho}
\end{equation} 
where we have used the relation between $\rho$ and interparticle
spacing $a$ in a hexagonal lattice, $\rho = 2/a^{2}\sqrt{3}$. The
constant $\alpha$ is calculated in reference \cite{ParoliniJCPM2015},
and is approximately equal to 0.0531.  The lower panel of
Fig.~\ref{fig:dr2VsRho} shows the results of applying this method to
10 crystalline samples at different areal densities. The error bars
show the discrepancy between the results of calculating the limiting
value of $\langle\delta \mathbf{r}_{\text{n.n.}}^{2}\rangle$ in two
different ways: first by using the plateau in the n.n.-MSD curve as a
function of time; and second by computing the variance of the
histogram of frame-to-frame displacements in the $x-$ and $y-$
directions separately. For a particle diffusing in an isotropic
harmonic potential, sampled over sufficiently long times, these two
methods should give the same result. Fitting the data to
Eq.~\ref{eq:drVsRho}, we find that $p = (455 \pm 20)\,$e$\cdot \mu$m.

When discussing phase behavior in this system, it is convenient to
introduce the dimensionless interaction parameter $\Gamma$,
\begin{equation*}
\Gamma = \frac{A(\pi \rho)^{3/2}}{\, k_{B}T},
\end{equation*}
where $T = 293\,$K is the temperature at which the experiments take
place. Using $\Gamma$ to describe the effective temperature of the
system allows us to directly compare our results with previous
experiments \cite{KusnerPRL1994, ZahnPRL1999} and simulations
\cite{LinPRE2006} using dipolar repulsive particles.

\begin{figure}[t!]
\center
\includegraphics[scale=0.37]{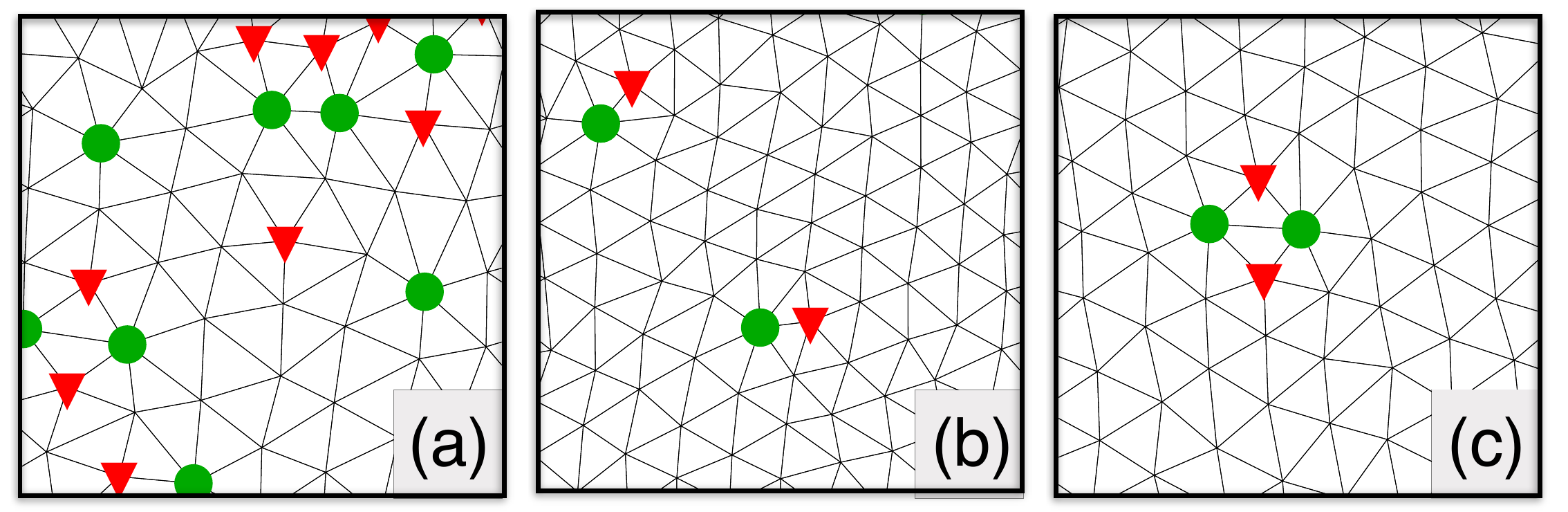}
\caption{(Color online.) Delaunay triangulations of typical particle
configurations, taken from the samples shown in
Fig.~\ref{fig:confocal_plus_trajectoires}. Five- and seven-coordinated
disclinations are marked by red triangles and green disks
respectively.  Sample (a) shows three unpaired disclinations, while
(b) shows two unpaired dislocations (5-7 pairs). Sample (c) shows a
pair of dislocations with opposite Burgers' vector. This configuration
is not a topological defect.
\label{topdefects}}
\end{figure}

\section{KTHNY theory: topological defects and orientational correlations}
Over the past several decades, KTHNY theory has been shown to describe
the phase behavior of a broad class of 2D materials, including dipolar
repulsive particles \cite{NelsonDefectsBook, GasserChemPhysChem2010,
KapferKrauthPRL2015}. According to this theory, melting of a 2D
crystalline solid takes place via two continuous transitions, which
can be understood in terms of the topological defects present in the
material. Two types of topological defects are important:
disclinations, points which have a number of nearest neighbors other
than six; and dislocations, bound pairs of one five-coordinated and
one seven-coordinated disclination. A dislocation is characterized by
a Burgers' vector, which represents the magnitude and direction of the
lattice distortion induced by the dislocation
\cite{HirthAndLothe}. Some examples of these kinds of defects are
shown in Fig.~\ref{topdefects} (a) and (b).

\begin{figure*}[t!]
\center
\includegraphics[scale=0.5]{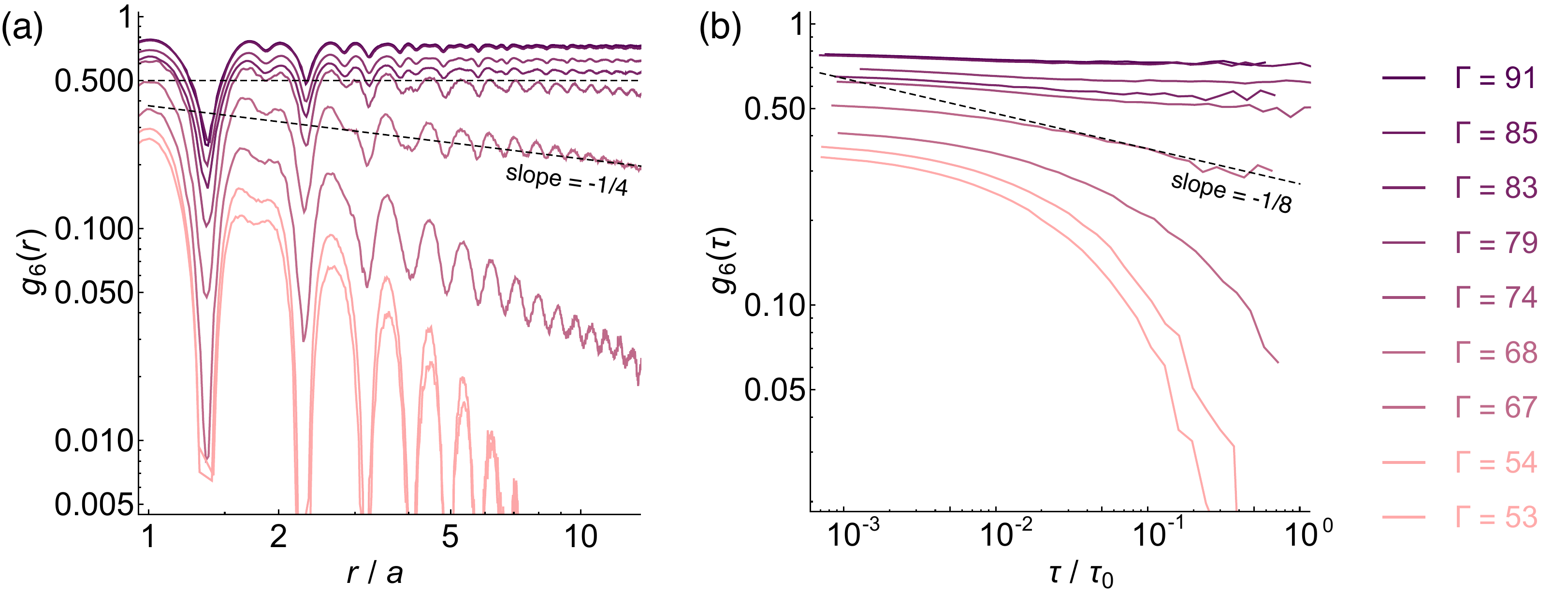}
\caption{(Color online.) Log-log plots of orientational correlation functions
$g_{6}(r)$ and $g_{6}(\tau)$, for nine samples at different values of
$\Gamma$. The sloped dashed lines in each plot indicate algebraic
decay with the exponents expected at the fluid-hexatic transition,
while the horizontal dashed line in the $g_{6}(r)$ plot separates the
curves which decay ($\Gamma \leq 74$) from those that reach a constant
value ($\Gamma \geq 79$). Using these curves, we identify the
fluid-hexatic and solid hexatic transitions at $\Gamma_{\textsc{FH}}
= 68 \pm 2$ and $\Gamma_{\textsc{HS}} = 76 \pm 3 $.
\label{fig:g6s}}
\end{figure*}

According to KTHNY theory, in the solid phase at equilibrium, no free
topological defects are present. However, there may be thermally
activated pairs of dislocations of opposite Burgers' vector which are
bound by an attractive potential. These kinds of structures, an
example of which is shown in Fig.~\ref{topdefects} (c), are not
topological, since they can occur via local rearrangements of the
lattice. At sufficiently low $\Gamma$, this attraction can be overcome
by thermal fluctuations, and the dislocation pairs
dissociate. Although the resulting free dislocations destroy the
finite shear modulus of the crystal lattice, the resulting material is
not an isotropic fluid, but rather a liquid crystalline hexatic
phase. The transition to the isotropic fluid is completed when the
dislocations themselves unbind into their constituent disclinations.

\begin {table}[b!]
\begin{center} 
\begin{tabular}{ | c || c | c | c  | } \hline 
\makecell{\phantom{m}\\\phantom{m} }   & \makecell{isotropic  \\ fluid}   & \makecell{\\ hexatic  \\ \phantom{.}} & solid  \\
\hline \hline  
\makecell{ \\\phantom{i} topological \phantom{i}\\ defects \\ \phantom{.}} &\makecell{ free \\ \phantom{i} disclinations\phantom{i}  } & \makecell{free \\ \phantom{i} dislocations \phantom{i} }  &  \makecell{none}    \\

\hline

\makecell{$g_{6}(r) \propto $ \\\tiny{$r \rightarrow \infty $}} & $\exp{(-r/\xi_{6})}$ &$r^{-\eta_{6}}$& \makecell{\\\phantom{M} const. \phantom{M}\\ \phantom{.}} \\

\hline

\makecell{$g_{6}(\tau) \propto $ \\\tiny{$\tau \rightarrow \infty $}} & $\exp{(-\tau/\tau_{6})}$&$\tau^{-\eta_{6}/2}$&   \makecell{\\const. \\ \phantom{.}} \\
\hline 

\end{tabular} 
\caption {Topological defect structure and properties of correlation
functions $g_{6}(r)$ and $g_{6}(\tau)$ in the solid, hexatic and fluid
phases, according to KTHNY theory. \label{tab:PhaseProperties}}
\end{center}
\end {table}

Using data from experiments and simulations, different groups have
explored various ways of quantitatively testing the predictions of
KTHNY theory \cite{DillmannJPCM2012, HanPRE2008}. Of particular
interest is the bond-orientational order parameter $\psi_{6}$, since
it can be easily calculated from real-space data, and because the
functional form of the associated correlation functions, $g_{6}(r)$ and
$g_{6}(\tau)$, clearly discriminate between the phases of the material.

If particle $k$ has position $\mathbf{r}_{k}$ at time $t$, the
bond-orientational order parameter $\psi_{6}(\mathbf{r}_{k},t)$ is the
defined by 
\begin{equation*}
\psi_{6}(\mathbf{r}_{k},t) = \frac{1}{n_{b}}\sum_{j  \, \in \, \mathcal{N}_{k}}e^{6i\theta_{kj}},
\end{equation*}
where the sum is taken over the $n_{b}$ nearest neighbors of particle
$k$. The angle between particle $k$ and its $j-$th neighbor,
$\theta_{kj}$, is taken with respect to an arbitrary but fixed
axis. The degree of local hexagonal order is given by $|\psi_{6}|$. In
the crystal phase, the orientation of the hexagonal 
unit cell is given by $\frac{1}{6}\arg{\psi_{6}}$. The space and time correlation functions
$g_{6}(r)$ and $g_{6}(\tau)$ are defined
\begin{equation*}
\begin{aligned}
g_{6}(r) &= \text{Re} \big{\{} \big{\langle} \psi_{6}^{*}(\mathbf{r}_{k},t) \psi_{6}(\mathbf{r}_{l},t)
\big{\rangle}_{\abs{\mathbf{r}_{k}-\mathbf{r}_{l}}=r} \big{\}} \quad \text{and} \\
g_{6}(\tau) &=  \text{Re} \big{\{} \big{\langle} \psi_{6}^{*}(\mathbf{r}_{k},t) \psi_{6}(\mathbf{r}_{k},t+\tau)
\big{\rangle}\big{\}}.
\end{aligned}
\end{equation*}
When calculating $g_{6}(r)$, the averages are taken over time and
pairs of particles $\{k,l\}$ satisfying the condition
$\abs{\mathbf{r}_{k}-\mathbf{r}_{l}}=r$. For $g_{6}(\tau)$, the
averages are taken over all particles $k$ and starting times
$t$. Thus, $g_{6}(r)$ is a two-particle correlation function, while
$g_{6}(\tau)$ is a single-particle quantity. According to KTHNY
theory, these correlation functions have distinct behaviors in each of
the three phases: for large $r$ and $\tau$, both functions tend to a
constant value in the solid phase, decay algebraically in the hexatic
phase, and decay exponentially in the isotropic fluid, with a
characteristic decay length [time] $\xi_{6}$ [$\tau_{6}$]. KTHNY
theory predicts that, in the hexatic phase, the exponent $\eta_{6}$ of
the power-law decay of $g_{6}(r)$ is twice the exponent of the
power-law decay of $g_{6}(\tau)$, and further dictates that, at the
fluid-hexatic transition, $\eta_{6}$ reaches a critical value of -1/4
\cite{NelsonDefectsBook}. Some of these predictions are summarized in
Table~\ref{tab:PhaseProperties}. 

Figure~\ref{fig:g6s} (a) and (b) respectively show $g_{6}(r)$ and
$g_{6}(\tau)$, plotted for nine samples at values of $\Gamma$ ranging
from 53 to 91. In our analysis of the temporal correlation function
$g_{6}(\tau)$, we rescale $\tau$ by the average time $\tau_{0}$
required for a freely diffusing particle to traverse the mean distance
between particles $a$. This step is necessary because the thickness of
the aqueous layer underneath the particles (see
Fig.~\ref{fig:conf_diag}) varies between samples. Thus, even in the
limit of very low particle density, different samples may have
diffusion coefficients $D_{0}$ that vary by as much as a factor of
four. We estimate $\tau_{0}$ from the small-time behavior of the
n.n.-MSD curves, such as those shown in Fig.~\ref{fig:dr2VsRho}. At
the smallest time intervals for which we have data, we assume that the
particle is freely diffusing inside its cage of nearest neighbors, and
fit the first two data points of the $\langle \delta
\mathbf{r}_{\text{n.n.}}(\tau)^{2} \rangle$ curve by a straight line,
containing the origin, with slope $4 D_{0}$ \cite{Note4}. This allows us to estimate the
time for a freely diffusing particle to traverse one interparticle
spacing, $\tau_{0} = a^{2}/4D_{0}$. For our samples, $\tau_{0}$ is
of order 30 mins.

For samples with $\Gamma \leq 67$, both $g_{6}(r)$ and $g_{6}(\tau)$ show the
exponential decay characteristic of the isotropic fluid phase, while
for $\Gamma \geq 79$, both functions tend to a constant
value at large $r$ and $\tau$.  For the sample at $\Gamma = 74$,
$g_{6}(r)$ shows behavior consistent with the power-law decay expected
in the hexatic phase. For $\Gamma = 68$, the slope of
$g_{6}(r)$ equals the critical value of -1/4 within experimental
error, and so we cannot unambiguously assign this sample to either phase.  As
can be seen from the right panel of Fig.~\ref{fig:g6s}, for the
samples at $\Gamma = 68$ and $\Gamma = 74$, our data for $g_{6}(\tau)$
do not allow us to distinguish between algebraic decay with a small
negative power, and a constant value.  Thus, we estimate that the
transition between the hexatic and the isotropic fluid phases occurs at
$\Gamma_{\textsc{FH}} = 68$, and the transition between the hexatic
and the ordered phase occurs at $\Gamma_{\textsc{HS}} = 76$.

We also note that the $g_{6}(r)$ and $g_{6}(\tau)$ curves for the
$\Gamma = 79$ sample lie above those for the sample at $\Gamma = 83$,
indicating that the former sample is more ordered. This apparent
non-monotonic behavior might reflect sample-to-sample variation in
electric dipole moment $p$, which would cause experimental uncertainty
in our calculated values of $\Gamma$, and may also account for some of
the spread of the data in Fig.~\ref{fig:dr2VsRho}. In fact, since the
limiting values of the correlation functions should be a monotonic
function of $\Gamma$, we can use the deviation from monotonicity to
estimate the uncertainty in our stated values of $\Gamma$. Doing this,
we find that the uncertainty in $\Gamma$ is approximately $3\%$. 
This figure only accounts for sample-to-sample
variation: the uncertainty in the mean value of the electric dipole
moment $p$ calculated from the data in Fig.~\ref{fig:dr2VsRho} could
lead to all the stated values of $\Gamma$ being shifted systematically
from their true values by as much as 8\%. In spite of these
experimental uncertainties, the values of $\Gamma_{\textsc{FH}}$ and
$\Gamma_{\textsc{HS}}$ that we find are in quantitative agreement
with those found in previous experiments \cite{ZahnPRL2000,
DeutschlanderPRL2014}.

\begin{figure}[t!]
\center
\includegraphics[scale=0.55]{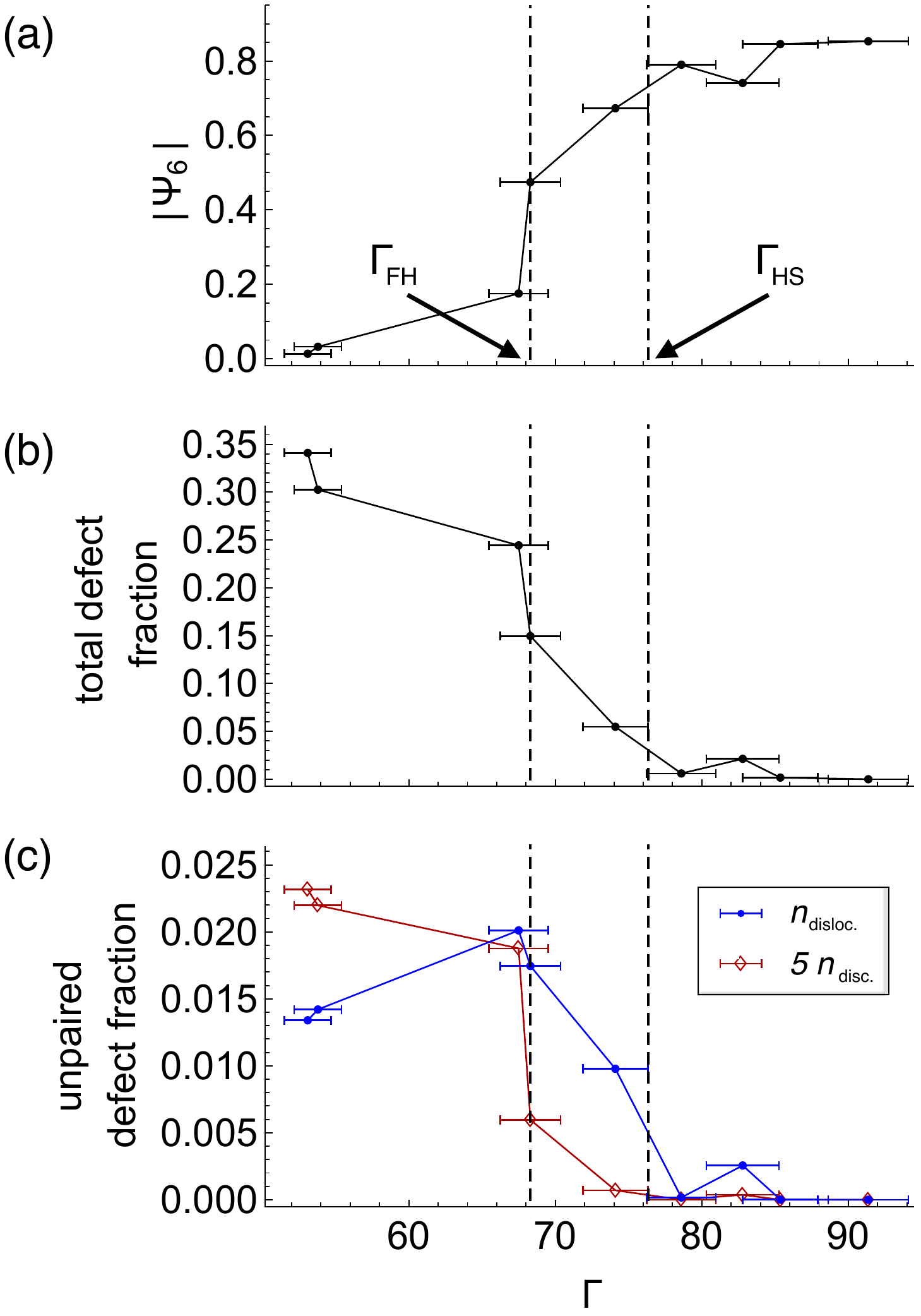}
\caption{(Color online.) (a) Absolute value of the 
average bond-orientational order parameter,
plotted as a function of interaction parameter $\Gamma$. The dashed
lines indicate the values of $\Gamma$ at the fluid-hexatic and
hexatic-solid transitions, as defined by the behavior of the
correlation functions $g_{6}(r)$ and $g_{6}(\tau)$ (see
Fig.~\ref{fig:g6s}). (b) Fraction of particles which do not
have six nearest neighbors. (c) Fraction of particles which
form unpaired disclinations or dislocations. To enable these
quantities to be displayed on the same plot, the fraction of unpaired
disclinations is multiplied by a factor of 5.
\label{fig:psirow}}
\end{figure}

The final orientational quantity we define is the average
orientational order parameter $\Psi_6 = \langle
\psi_{6}(\mathbf{r}_{k},t) \rangle$, where the average is taken over all
points $k$ and times $t$. We expect that, in the isotropic fluid
phase, $|\Psi_6| = 0$, while in the crystalline solid phase,
$|\Psi_6|$ takes a finite positive value, which tends to unity as
$\Gamma$ increases. In the top panel of Fig.~\ref{fig:psirow}, we plot
$|\Psi_{6}|$ as a function of interaction parameter $\Gamma$. The middle
panel of the same figure shows the total defect fraction, defined as
the fraction of particles which have a number of nearest neighbors
other than six. Between $\Gamma_{\textsc{FH}}$ and
$\Gamma_{\textsc{HS}}$, the defect fraction drops dramatically, and we
see a corresponding growth in the orientational order $|\Psi_{6}|$ in
the system.

The bottom panel of Fig.~\ref{fig:psirow} shows a plot of the fraction
of unpaired dislocations and disclinations, $n_{\text{disloc.}}$ and
$n_{\text{disc.}}$. A topological defect is defined as unpaired if it
is does not share a Delaunay bond with any other defect. Thus, the
5-coordinated disclination in the center of Fig.~\ref{topdefects} (a)
is unpaired, as are the two dislocation defects in
Fig.~\ref{topdefects} (b). On the other hand, both the cluster of
defects in the top right of Fig.~\ref{topdefects} (a) and the two
adjacent dislocations of opposite Burgers' vector in
Fig.~\ref{topdefects} (c) are paired, and neither contributes to
$n_{\text{disloc.}}$ or $n_{\text{disc.}}$.  While computationally
straightforward, our definition of \textit{unpaired} defects does not
provide a direct measurement of the concentration of \textit{free}
defects in the sense of KTHNY theory. This is true for at least two
reasons: first, our definitions of $n_{\text{disloc.}}$ and
$n_{\text{disc.}}$ treat energetically bound but non-adjacent defects
as unpaired. This leads us to overestimate the number of free defects,
especially in the solid and hexatic phases. Second, our definitions
completely neglect defect clusters, such as the structure in the top
right of Fig.~\ref{topdefects} (a), which may contain one or several
net topological defects. This will cause us to underestimate the
number of free defects, especially in the fluid phase, where such
clusters proliferate. Despite these shortcomings, we find that
$n_{\text{disloc.}}$ and $n_{\text{disc.}}$ display the expected
behavior in the vicinity of the transitions: near $\Gamma =
\Gamma_{\textsc{FH}} $, $n_{\text{disc.}}$ drops dramatically, while
near $\Gamma = \Gamma_{\textsc{HS}} $, $n_{\text{disloc.}}$ does the
same.

\section{Dynamical measures of phase behavior \label{sec:dynamics}}
\begin{figure}[t!]
\center
\includegraphics[scale=0.5]{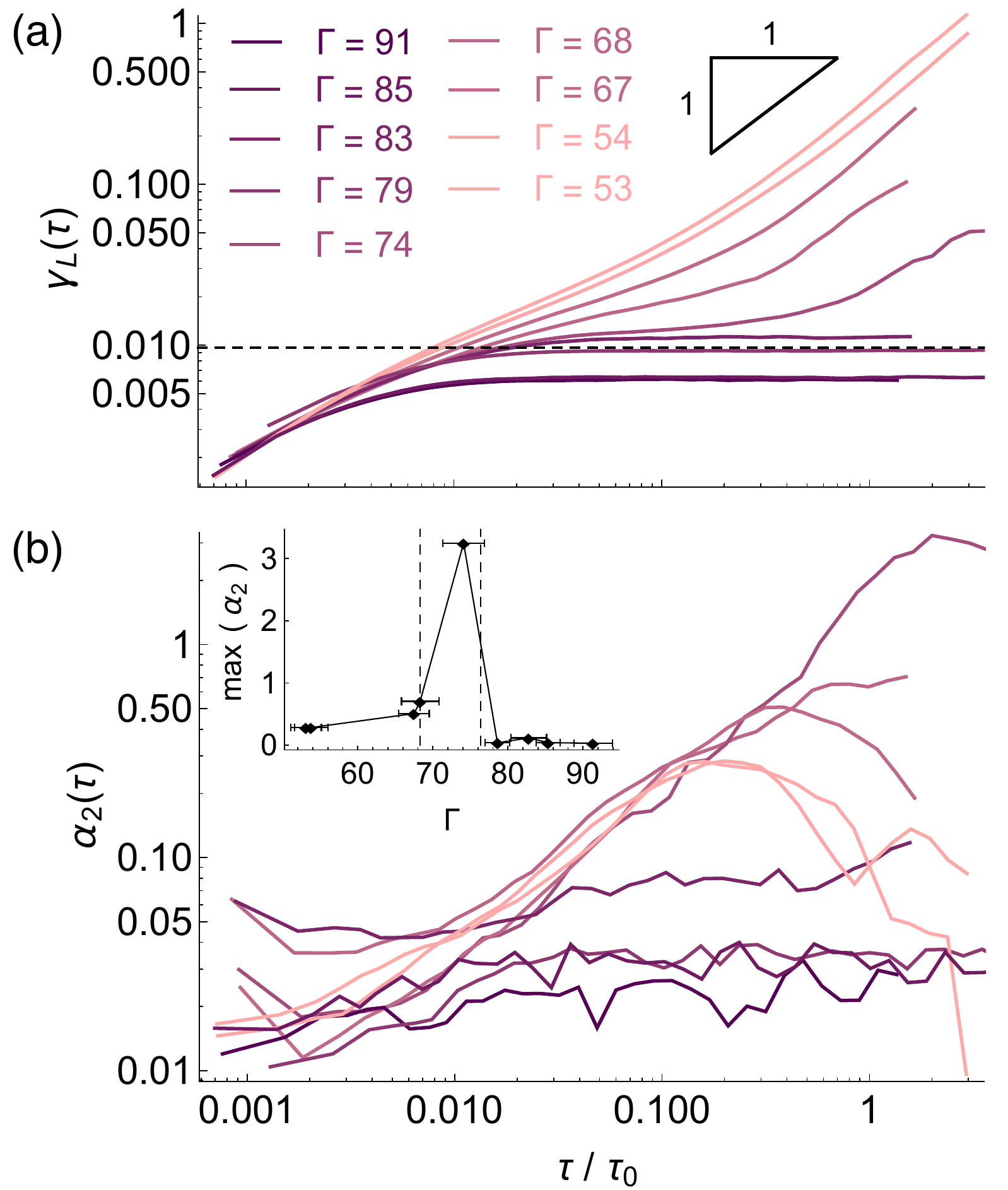}
\caption{(Color online.) Log-log plots of: (a) dynamic Lindemann parameter
$\gamma_{\textsc{L}}(\tau)$; and (b) non-Gaussian parameter
$\alpha_{2}(\tau)$. Panel (a) shows that, for samples identified as belonging
to the ordered phase by our analysis of the correlation functions,
$\gamma_{\textsc{L}}(\tau)$ reaches a constant value, while the
samples with $\Gamma < \Gamma_{\textsc{HS}}$ behave diffusively or
subdiffusively. The horizontal dashed line indicates the critical
value $\gamma_{L}^{c} = 0.0097$ predicted by Eqns.~\eqref{eq:drVsRho}
and \eqref{LindemannParam}, evaluated at the density
$\rho_{\textsc{HS}}$ corresponding to $\Gamma_{\textsc{HS}}$.  Panel
(b) shows that, in the isotropic fluid, $\alpha_{2}$ peaks at
some intermediate timescale, while in the candidate hexatic samples
($\Gamma =$ 68 and 74), non-Gaussian behavior grows over the
timescales we observe. The inset in the lower plot highlights the
dramatic growth in the maximum value of $\alpha_{2}$ in the hexatic
phase. The vertical dashed lines indicate $\Gamma_{\textsc{FH}}$, and
$\Gamma_{\textsc{HS}}$, as defined by the behavior of the correlation
functions $g_{6}(r)$ and $g_{6}(\tau)$.
  \label{fig:dynamicQuants}}
\end{figure}

As well as displaying distinctive spatial structure, different phases
of a material are typically characterized by their dynamics.  To
investigate this aspect of the phase behavior of our samples, we plot,
in Fig.~\ref{fig:dynamicQuants}, the dynamic Lindemann parameter
$\gamma_{L}(\tau)$ and the non-Gaussian parameter $\alpha_{2}(\tau)$
\cite{ZahnPRL2000}. The dynamic Lindemann parameter $\gamma_{L}(\tau)$
is defined
\begin{equation}
\gamma_{L}(\tau) = \frac{\langle \delta
  \mathbf{r}_{\text{n.n.}}(\tau)^{2} \rangle }{2 a^{2}},
\label{LindemannParam}
\end{equation}
where the n.n.-MSD $\langle \delta \mathbf{r}_{\text{n.n.}}(\tau)^{2} \rangle$ is
defined in Eqn.~\eqref{eq:nnMSd}. The non-Gaussian parameter
$\alpha_{2}(\tau)$ is defined as
\begin{equation*}
\alpha_{2}(\tau) = \frac{\langle \delta
  \mathbf{r}_{\text{n.n.}}(\tau)^{4}\rangle}{2 \langle \delta
  \mathbf{r}_{\text{n.n.}}(\tau)^{2}\rangle^{2}}-1.
\end{equation*}
This quantity measures the extent to which the histogram of particle
displacements deviates from the normal distribution. Using the phase
classification based on the behavior of the correlation functions
$g_{6}(r)$ and $g_{6}(\tau)$, our measurements of $\gamma_{L}(\tau)$
and $\alpha_{2}(\tau)$ are consistent with previous measurements of
the dynamics of systems in the solid, fluid and hexatic
phases~\cite{ZahnPRL2000}.

In the solid phase, the dynamic Lindemann parameter reaches a plateau
value at long time. The observed critical value of the dynamic
Lindemann parameter is $\gamma_{L}^{c} = 0.012 \pm 0.001$. This is
close to the value predicted by evaluating Eqns.~\eqref{eq:drVsRho}
and~\eqref{LindemannParam} at the density $\rho_{\textsc{HS}}$
corresponding to the hexatic-solid transition, $\gamma_{L}^{c} =
0.0097\pm 0.005$. This comparison serves as a consistency check on
Eqn.~\eqref{eq:drVsRho}, and the measured value of
$\Gamma_{\textsc{HS}}$. In the solid phase, the non-Gaussian parameter
$\alpha_{2}(\tau)$ has a small positive value independent of time.
This may be because the quantities $\mathbf{r}_{i,
\text{n.n.}}$ are not statistically independent, as they take into
account the positions of the nearest neighbors.

In the fluid phase, we observe diffusive behavior, $\gamma_{L}(\tau)
\propto \tau$, at long times, while the non-Gaussian parameter
$\alpha_{2}(\tau)$ displays a local maximum at time intervals
$\tau/\tau_{0} \sim 0.1$. These timescales also correspond to the
presence of a shoulder in the $\gamma_{L}(\tau)$ curves, and are also
similar to the characteristic times $\tau_{6}$ of the exponential
decay of the $g_{6}(\tau)$ curves shown in Fig.~\ref{fig:g6s}. All
these timescales may originate in collective rearrangements of defect
clusters, such as that shown in Fig.~\ref{topdefects} (a).

In the sample we identify as belonging unambiguously to the hexatic
phase ($\Gamma = 74$), $\gamma_{L}(\tau)$ behaves subdiffusively over
observed times.  For the sample at $\Gamma = 68$, the correlation
functions $g_{6}(r)$ and $g_{6}(\tau)$ are consistent with critical
behavior, and we are unable to assign it to either the fluid or the
hexatic phase. For this sample, the slope of $\gamma_{L}(\tau)$
appears to be approaching 1 at the longest times we measure, perhaps
indicating that it is indeed a fluid. Previously, Zahn and Maret
showed that $\alpha_{2}(\tau)$ tends to a constant value of order
unity for systems in the hexatic phase~\cite{ZahnPRL2000}. Our data
are consistent with this finding, but we do not record our candidate
hexatic samples for sufficiently long times to verify the limiting
behavior. We do, however, observe the sharp growth in the maximum
value of $\alpha_{2}(\tau)$ in the hexatic phase that was reported in
the same study.

\section{Conclusions} 
In this work, we study a system of charged colloidal particles that
are electrostatically bound to a fluid interface, and interact via electric
dipole-dipole repulsion. We show that the phase behavior of this
system is well-described by KTHNY theory, with density
as the control parameter. Using the orientational correlation
functions $g_{6}(r)$ and $g_{6}(\tau)$, we assign each sample to the
solid, isotropic fluid, or hexatic phase. We demonstrate that the concentration
of unpaired dislocations and disclinations are consistent with the
KTHNY picture of defect-mediated melting. Finally, we find that each
phase displays distinctive dynamical behavior, as measured by the the
dynamic Lindemann parameter $\gamma_{L}(\tau)$ and the non-Gaussian
parameter $\alpha_{2}(\tau)$.

\begin{figure}[b!]
\center
\includegraphics[scale=0.28]{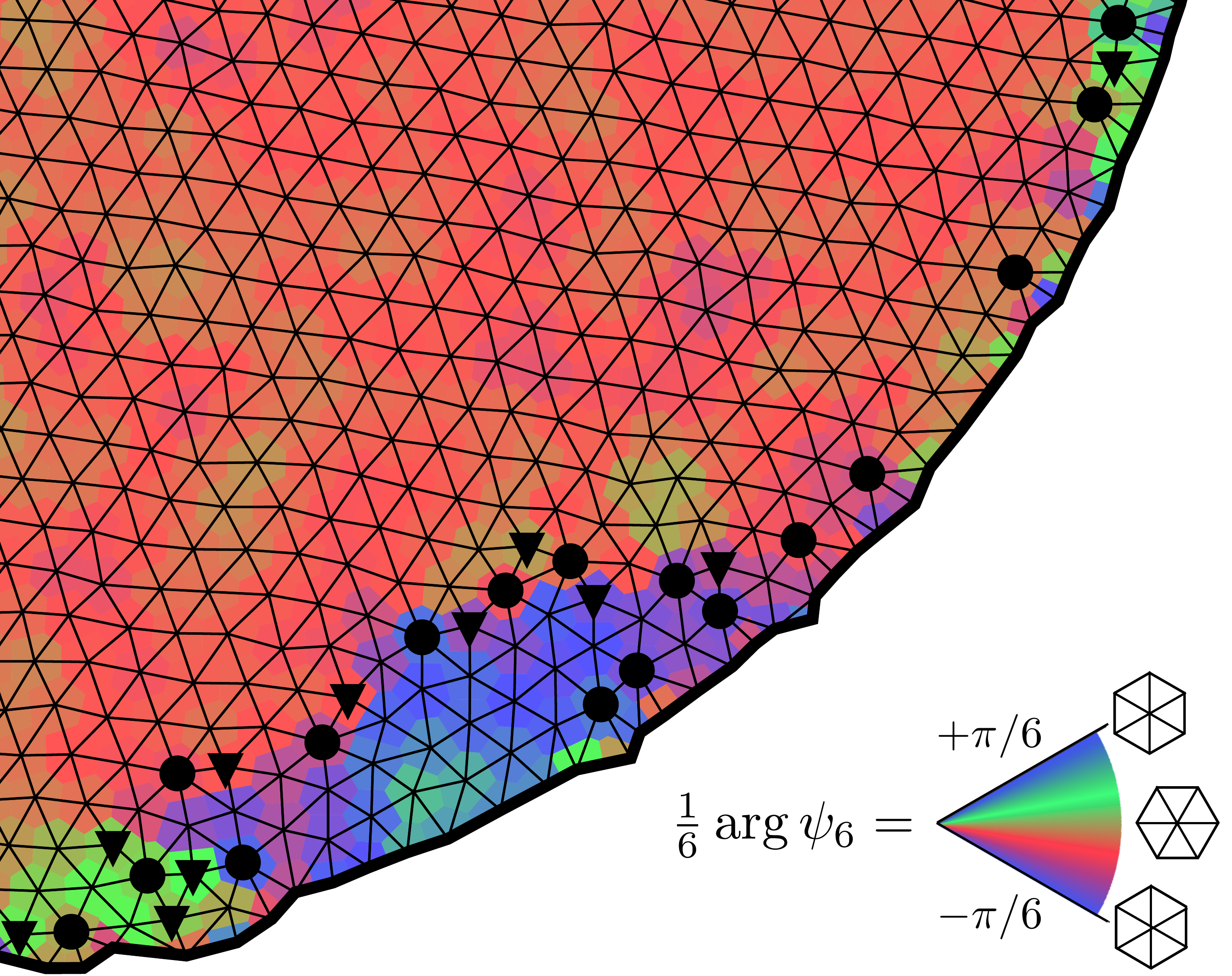}
\caption{(Color online.) Plot of the Delaunay triangulation of
interfacially bound particles in a single quadrant of a roughly
circular droplet. This sample has $\Gamma = 120$, and approximately
2000 particles in total. The color map shows the local orientation of
the lattice, given by $\frac{1}{6}\arg{\psi_{6}}$, while 5- and
7-coordinated disclinations are indicated by triangles and disks
respectively. The thick black curve identifies the droplet
edge. (Disclinations are not plotted for the particles on the
boundary.) Away from the edge, the system forms a monocrystal that
spans the interior of the droplet. Immediately adjacent to the droplet
edge, the lattice is aligned with the edge. Where the orientation of
the interior region does not match that of the edge, grain boundaries
form, a few lattice spacings from the boundary. These grain boundaries
are delineated by a chain of polarized dislocations (5-7 pairs).
  \label{fig:edgeGrain}}
\end{figure}

Given the small number of particles in our system, the extent to which
our data are well-modeled by KTHNY theory is quite surprising. This
agreement is only possible because the orientation of the droplet
edge, which is delineated by a line of pinned particles (see
Fig.~\ref{fig:confocal_plus_trajectoires}), does not propagate into
the interior of the droplet. In the fluid phase, this is expected,
since orientational correlations decay exponentially over lengths of a
few interparticle spacings -- this decay is evident in the $\Gamma
\leq 67$ curves in Fig.~\ref{fig:g6s} (a). In the crystal phase, as
shown in Fig.~\ref{fig:edgeGrain}, the orientation inherited from the
droplet edge is destroyed by a series of grain boundaries that run
around the inner perimeter of the droplet. These grain boundaries
separate an interior monocrystalline region from an outside layer, a
few interparticle spacings wide, which is aligned with the edge of the
droplet. Evidently, the strain fields associated with these grain
boundaries are not large enough to significantly disrupt the phase
behavior of the material in the interior region. In the hexatic, it is
not immediately clear how the system accommodates the presence of the
droplet edge, since grain boundaries are difficult to identify
unambiguously in this phase.

The concurrence between our findings and previous work on larger
systems of dipolar repulsive particles extends to quantitative
agreement on the transition values of the interaction parameter
$\Gamma$ \cite{ZahnPRL1999, DeutschlanderPRL2014}, although the
sparsity our data, as well as the error bars for $\Gamma$ shown for
example in Fig.~\ref{fig:psirow}, limit our ability to determine the
width of the hexatic window.

This work was supported primarily by the National Science Foundation
under Award No. DMR-1105417, and partially by the Materials Research
Science and Engineering Center (MRSEC) program of the National Science
Foundation under Award No. DMR-1420073. Additional financial support 
was provided by NASA (NNX13AR67G).


\end{document}